\documentclass[shortnote,twocolumn]{jpsj3}
\usepackage{braket}

\title{One-Dimensional Integrable Spinor BECs Mapped to Matrix Nonlinear Schr\"odinger Equation and Solution of Bogoliubov Equation in These Systems}

\author{Daisuke A. Takahashi\thanks{E-mail address: takahashi@vortex.c.u-tokyo.ac.jp} 
}
\inst{Department of Basic Science, the University of Tokyo, Tokyo 153-8902 
}
\kword{matrix nonlinear Schr\"odinger equation, spinor Bose-Einstein condensate, Bogoliubov equation, squared Jost function.}

\begin{document}
\maketitle
	Nonlinear Schr\"odinger equation(NLSE) is one of the oldest integrable nonlinear equations, which was solved by the inverse scattering method(ISM)\cite{ZakharovShabat}. In the context of condensed matter physics, NLSE is also called Gross-Pitaevskii equation, and it describes the dynamics of one-dimensional(1D) Bose-Einstein condensate(BEC). Recently, integrable 1D BECs with internal spin degree of freedom (spinor BECs) have been discovered\cite{IedaMiyakawaWadati,GerdjikovKostovValchev}. First, the spin-1 integrable system has been discovered\cite{IedaMiyakawaWadati} by finding a mapping to the matrix NLSE(MNLSE)\cite{TsuchidaWadati}:
	\begin{align}
		\mathrm{i}\partial_t Q=-\partial_x^2 Q\pm 2QQ^\dagger Q. \label{eq:mnlse}
	\end{align}
	Here $ Q $ is a square matrix, and the minus(plus) sign corresponds to the self-(de)focusing case. In the context of BEC, the minus(plus) sign represents the system of attractive(repulsive) bosons. Subsequently, integrable BEC systems are found for every integer spin-$ n $\cite{GerdjikovKostovValchev}. The energy functional of integrable spinor BECs are given by\cite{GerdjikovKostovValchev}
	\begin{align}
		H = \int\!\!\mathrm{d}x \biggl[\, \sum_{m=-n}^n \left| \frac{\partial \psi_m}{\partial x} \right|^2 \pm\frac{1}{2^{n-1}}\Bigl( \rho^2-\frac{1}{2}|\Theta|^2\Bigr) \biggr], \label{eq:funcspinor}
	\end{align}
	where  $ \psi_m $ is a condensate wavefunction with magnetic quantum number $ m $, $ \rho=\sum_{m=-n}^n |\psi_m|^2 $ is a number density, and $ \Theta=\sum_{m=-n}^n (-1)^m\psi_m\psi_{-m} $ is a singlet pair amplitude. The coefficient $ 2^{n-1} $ can be eliminated by a choice of normalization, but we keep it for later convenience.\\
	\indent In this short note, we show that the above integrable systems (\ref{eq:funcspinor}) are all mapped to MNLSE, and solve the Bogoliubov equation of these systems. While the integrability itself has been already proved by using a different Lax pair\cite{GerdjikovKostovValchev}, the mapping to MNLSE has the following advantages: (i) The explicit expression for $ N $-soliton solution is already given under both vanishing\cite{TsuchidaWadati} and non-vanishing\cite{IedaUchiyamaWadati} boundary conditions. (ii) With the aid of the theory of squared eigenfunction(or squared Jost function)\cite{Kaup,ChenChenHuang}, solutions of Bogoliubov equation can be obtained. Since there exist various kinds of low-energy excitations in spinor BEC systems\cite{TLHospin1}, it would be an interesting future work to solve the scattering problem of low-energy excitations in the presence of an external potential\cite{KovrizhinDanshita} by using the solution given in this short note. \\
	\indent \textit{Mapping to MNLSE---} We first write the mapping matrix and its (anti)symmetrization, and next explain how to find it. The mapping from spin-$ n $ integrable spinor BEC to  $ 2^n $-dimensional MNLSE is given by
	\begin{align}
	\begin{split}
		Q &= \frac{1}{2^{n/2}}\biggl( \sum_{m=1}^n \Bigl[ \psi_m(-1)^m I^{\otimes(n-m)}\!\otimes\! \sigma_+\!\otimes\!\sigma_z^{\otimes(m-1)} \\ 
		&\qquad+\psi_{-m} I^{\otimes(n-m)}\!\!\otimes\! \sigma_-\!\otimes\!\sigma_z^{\otimes(m-1)}\Bigr]+\psi_0\sigma_z^{\otimes n}\biggr). 
	\end{split}\label{eq:mapQ}
	\end{align}
	Here $ \sigma_i(i=x,y,z) $ are Pauli matrices, $ \sigma_{\pm}=\frac{1}{\sqrt{2}}(\sigma_x\pm\mathrm{i}\sigma_y) $,  $ I $ is $ 2\times2 $ identity matrix, and $ A^{\otimes i}:=\underbrace{A\otimes\dotsb\otimes A}_{i \text{ times}}$. (for $ i=0 $, we define $ A^{\otimes 0}:=1 $.) The energy functional which yields Eq. (\ref{eq:mnlse}) is 
	\begin{align}
		H=\int\!\!\mathrm{d}x\left[ \operatorname{tr}\left( \frac{\partial Q^\dagger}{\partial x}\frac{\partial Q}{\partial x} \right)\pm\operatorname{tr}(Q^\dagger QQ^\dagger Q) \right].
	\end{align}
	Substituting the matrix (\ref{eq:mapQ}) to this functional, one can obtain the functional (\ref{eq:funcspinor}). \\ 
	\indent Furthermore, we can (anti)symmetrize the above matrix by the unitary matrix
	\begin{gather}
		\mathcal{V}=\begin{cases} (\mathrm{i}\sigma_y\otimes\sigma_x)^{\otimes n/2} & (n: \text{even}) \\ (\mathrm{i}\sigma_y\otimes\sigma_x)^{\otimes (n-1)/2}\otimes(\mathrm{i}\sigma_y) &(n: \text{odd}) \end{cases}.
	\end{gather}
	The matrix $ \tilde{Q}=Q\mathcal{V} $ then satisfies  $ \tilde{Q}^T = (-1)^{\frac{1}{2}n(n-1)}\tilde{Q} $.\\ 
		\indent We briefly summarize how to construct the matrix (\ref{eq:mapQ}). In the theory of Lie algebra\cite{Georgi}, writing the elements of Cartan subalgebra as $ H_1,\dots,H_r $ ($ r $: rank of this algebra) and raising or lowering operators as $ E_\alpha $(which changes weight from $ \mu $ to $ \mu+\alpha $), an irreducible representation $ D $ is determined from the following relations:
	\begin{align}
		H_i \ket{\mu,D} &= \mu_i \ket{\mu,D}, \\
		E_\alpha \ket{\mu,D} &= N_{\alpha,\mu}\ket{\mu+\alpha,D}.
	\end{align}
	Here $ \mu=(\mu_1,\dots,\mu_r) $ is a weight vector, and $ N_{\alpha,\mu} $ is a representation-dependent constant. The irreducible tensor operator\cite{Iachello} corresponding to this representation, $ D $, is characterized by the following commutation relations:
	\begin{align}
		[H_i,T_{\mu,D}] &= \mu_i \,T_{\mu,D}\,, \label{eq:comto1}\\
		[E_\alpha,T_{\mu,D}] &= N_{\alpha,\mu}T_{\mu+\alpha,D}\,. \label{eq:comto2} 
	\end{align}
	\indent Now, let us consider the algebra $ so(2n\!+\!1) $. We construct the tensor operators corresponding to  $ 2n\!+\!1 $-dimensional fundamental representation (we write it as $ \boldsymbol{2n\!+\!1} $ --- it is unitary equivalent to the definition of the algebra itself.). As for the matrix representation of generators, $ H_i $s and $ E_\alpha $s, we use the $ 2^n $-dimensional spinor representation. Therefore tensor operators also become $ 2^n $-dimensional matrices.  In this representation, generators are represented by the $ n $-fold tensor product of Pauli matrices.\cite{Georgi} One can determine the matrix elements of $ T_{\mu,\boldsymbol{2n+1}} \ (\mu=0,\pm e_1,\dots,\pm e_n)$ through the commutation relations (\ref{eq:comto1}) and (\ref{eq:comto2}), and obtain the matrices appearing in Eq. (\ref{eq:mapQ}).\\ 
	\indent \textit{Bogoliubov equation --- }Next, we solve the Bogoliubov equation of these integrable systems. Since we are interested in an application to the scattering problem of collective excitations\cite{KovrizhinDanshita}, we discuss the problem of non-vanishing boundary condition\cite{IedaUchiyamaWadati}, and consider MNLSE of the self-defocusing case with a chemical potential term: 
	\begin{align}
		\mathrm{i}\partial_t Q = -(\partial_x^2+\mu)Q+2QQ^\dagger Q. \label{eq:mnlsewithmu}
	\end{align}
	The Bogoliubov equation, though its standard derivation is diagonalization of second-quantized Hamiltonian in mean field approximation, is easily obtained by substituting $ Q=Q+\delta Q $ in Eq. (\ref{eq:mnlsewithmu}) and discarding the higher order terms of $ \delta Q $. Rewriting $ (\delta Q,-\delta Q^\dagger)=(U,V)$, one obtains
	\begin{align}
	\begin{split}
		\mathrm{i}\partial_t U&=-(\partial_x^2+\mu)U+2\bigl(QQ^\dagger U \!+\! UQ^\dagger Q \!-\! QVQ\bigr), \\
		\mathrm{i}\partial_t V&=(\partial_x^2+\mu)V-2\bigl(Q^\dagger QV \!+\! VQQ^\dagger \!-\! Q^\dagger UQ^\dagger\bigr).
	\end{split} \label{eq:Bogoliubov}
	\end{align}
	ISM of MNLSE is formulated through the following extended Zakharov-Shabat type eigenvalue problem\cite{TsuchidaWadati,IedaUchiyamaWadati}:
	\begin{align}
		&\frac{\partial }{\partial x}\!\begin{pmatrix} \vec{f} \\ \vec{g} \end{pmatrix}\! = \!\begin{pmatrix} -\mathrm{i}\lambda & Q \\ Q^\dagger & \mathrm{i}\lambda \end{pmatrix}\!\begin{pmatrix} \vec{f} \\ \vec{g} \end{pmatrix}\!, \label{eq:ZS01}\\
		&\frac{\partial }{\partial t}\!\begin{pmatrix} \vec{f} \\ \vec{g} \end{pmatrix}\! = \!\begin{pmatrix} \mathrm{i}(-2\lambda^2\!+\!\frac{1}{2}\mu\!-\!QQ^\dagger) \!\!& 2\lambda Q+\mathrm{i}\partial_xQ \\ 2\lambda Q^\dagger-\mathrm{i}\partial_xQ^\dagger \!\!& \mathrm{i}(2\lambda^2\!-\!\frac{1}{2}\mu\!+\!Q^\dagger Q) \end{pmatrix}\!\!\begin{pmatrix} \vec{f} \\ \vec{g} \end{pmatrix}\!. \label{eq:ZS02}
	\end{align}
	Here $ \lambda $ is a spectral parameter, and $ \vec{f} $ and $ \vec{g} $, which are called Jost function, are $ n $-component vectors when $ Q $ is an $ n\times n $ matrix. The compatibility condition $ \partial_x\partial_t(\vec{f},\vec{g})^T=\partial_t\partial_x(\vec{f},\vec{g})^T $ reproduces Eq. (\ref{eq:mnlsewithmu}).\\
	\indent Assume that $ Q $ is (anti)symmetric: $ Q=\varepsilon Q^T \ (\varepsilon=\pm1)$. We can then show by a straightforward calculation that if $ (\vec{f}_1,\vec{g}_1)^T $ and $ (\vec{f}_2,\vec{g}_2)^T $ are the solutions of Eqs. (\ref{eq:ZS01}) and (\ref{eq:ZS02}) with the same $ \lambda $, $ (U,V)=(\vec{f}_1\vec{f}_2^T,-\varepsilon \vec{g}_1\vec{g}_2^T) $ satisfies Eqs. (\ref{eq:Bogoliubov}). Thus, the squared Jost function gives a solution of Bogoliubov equation. \\ 
	\indent \textit{Example --- }As an example, let us consider the spin-1 case in the presence of one dark soliton\cite{UchiyamaIedaWadati}. We want the stationary solution of Bogoliubov equation with eigenenergy $ \epsilon $, which is obtained by the replacement $ \mathrm{i}\partial_t\rightarrow\epsilon $ in Eqs. (\ref{eq:Bogoliubov}). Correspondingly, we must consider Eqs. (\ref{eq:ZS01}) and (\ref{eq:ZS02}) with $ \mathrm{i}\partial_t\rightarrow \epsilon/2 $, since a quadratic form of $ \vec{f} $ and $ \vec{g} $ gives a solution of Bogoliubov equation.\\
	\indent In order to simplify the mathematical descriptions, we first prepare several notations for the solutions of one-component NLSE (i.e., a scalar BEC). The one dark soliton solution in the comoving frame is 
	\begin{gather}
		\phi_{\text{s}}(x;x_0)=-\mathrm{e}^{\mathrm{i}\varphi}\mathrm{e}^{\mathrm{i}px}\left[ p+\mathrm{i}q\tanh(q(x\!-\!x_0)) \right],
	\end{gather}
	where  $ p=-\lambda_0\cos\varphi, q=\lambda_0\sin\varphi,\ \lambda_0>0 $ and $ \varphi\in(0,\pi) $. The subscript ``s'' means scalar. For this solution, the chemical potential becomes $\mu=p^2+2\lambda_0^2$. We also define $ \phi_{\text{s}}(x;-\infty):=\lambda_0\mathrm{e}^{\mathrm{i}px} $. The solution of Bogoliubov equation is given by $ (u,v)=(f^2,-g^2) $. Considering the solutions in the uniform system, one can show that the wavenumber of an excitation $ k $ and the spectral parameter $ \lambda $ are related as $ k=-\epsilon/(2\lambda-p) $. The expression for $ (f,g) $ in the presence of one dark soliton is given by\cite{ChenChenHuang}
	\begin{align}
		&\begin{pmatrix} f_{\text{s}}(x;x_0) \\ g_{\text{s}}(x;x_0) \end{pmatrix} = \mathrm{e}^{\mathrm{i}kx/2}\times \\
		& \quad \begin{pmatrix} \mathrm{e}^{\mathrm{i}(px+\varphi)/2} \bigl[  \mathrm{i}q\tanh(q(x\!-\!x_0))+(k/2)+(\epsilon/2k) \bigr] \\ -\mathrm{i}\mathrm{e}^{-\mathrm{i}(px+\varphi)/2} \bigl[ \mathrm{i}q\tanh(q(x\!-\!x_0))+(k/2)-(\epsilon/2k) \bigr] \end{pmatrix}\!, \nonumber
	\end{align}
	where the wavenumber $ k $ satisfies the dispersion relation $ (\epsilon-2kp)^2=k^2(k^2+4\lambda_0^2) $. We also define $ f_{\text{s}}(x;-\infty):= \mathrm{e}^{\mathrm{i}[(k+p)x+\varphi]/2} \left[ \mathrm{i}q+(k/2)+(\epsilon/2k) \right] $ and $ g_{\text{s}}(x;-\infty):=-\mathrm{i} \mathrm{e}^{\mathrm{i}[(k-p)x-\varphi]/2} \left[ \mathrm{i}q+(k/2)-(\epsilon/2k) \right] $. \\
	\indent Having prepared the notations, let us move on to the spin-1 system. The symmetrized matrix is given by $ Q=\left(\begin{smallmatrix}\psi_1&\psi_0/\sqrt{2}\\ \psi_0/\sqrt{2}&\psi_{-1}\end{smallmatrix}\right) $, and correspondingly, Bogoliubov wavefunctions become $ U=\left(\begin{smallmatrix}u_1&u_0/\sqrt{2}\\ u_0/\sqrt{2}&u_{-1}\end{smallmatrix}\right) $ and  $ V=\left(\begin{smallmatrix}v_1&v_0/\sqrt{2}\\ v_0/\sqrt{2}&v_{-1}\end{smallmatrix}\right) $. The one dark soliton solution\cite{UchiyamaIedaWadati} can be diagonalized as
	\begin{align}
		Q = D \begin{pmatrix}\phi_{\text{s}}(x;x_1) & 0 \\ 0 & \phi_{\text{s}}(x;x_2) \end{pmatrix}  D^{T}. \label{eq:spin1ds}
	\end{align}
	Here $ D $ is a rotation matrix of spin-1/2. (Note that the rightmost term is not $ D^\dagger $ but $ D^T $\cite{UchiyamaIedaWadati}.) If either $ x_1 $ or $ x_2 $ equals to $ -\infty $, Eq. (\ref{eq:spin1ds}) represents a ferromagnetic soliton\cite{UchiyamaIedaWadati}. If both are finite, it represents a polar soliton.\cite{UchiyamaIedaWadati} Jost functions are given by
	\begin{align}
		\begin{pmatrix}\vec{f} \\ \vec{g} \end{pmatrix}=\begin{pmatrix} f_{\text{s}}(x;x_i) D\vec{e}_i \\  g_{\text{s}}(x;x_i) D^* \vec{e}_i \end{pmatrix}, \quad (i=1,2),
	\end{align}
	where $ \vec{e}_1=(1,0)^T $ and $ \vec{e}_2=(0,1)^T $. From these Jost functions, we can construct the solutions of the Bogoliubov equation as
	\begin{align}
	\begin{split}
		\begin{pmatrix}U \\ V \end{pmatrix} =& \begin{pmatrix} f_{\text{s}}(x;x_1)^2 D (I+\sigma_z) D^T \\  -g_{\text{s}}(x;x_1)^2 D^* (I+\sigma_z) D^\dagger \end{pmatrix}, \\
		& \begin{pmatrix} f_{\text{s}}(x;x_2)^2 D (I-\sigma_z) D^T \\  -g_{\text{s}}(x;x_2)^2 D^* (I-\sigma_z) D^\dagger \end{pmatrix}, \\
		& \begin{pmatrix} f_{\text{s}}(x;x_1)f_{\text{s}}(x;x_2) D \sigma_x D^T \\  -g_{\text{s}}(x;x_1)g_{\text{s}}(x;x_2) D^* \sigma_x D^\dagger \end{pmatrix}.
	\end{split}\label{eq:bogospin1sol}
	\end{align}
	In the integrable case, the dispersion relations of excitations of the spin fluctuation and the density fluctuation are degenerate. (See Ref. \citen{TLHospin1} for the general case; Note that the asymptotic form of Eq. (\ref{eq:spin1ds}) is always polar state, regardless of whether the soliton is ferromagnetic or polar.) Therefore any linear combination of the above Eq. (\ref{eq:bogospin1sol}) becomes the solution. However, if one wants to separate the spin fluctuation part and the density fluctuation part, one needs to make a special linear combination of them. \\
	\indent Note added: After submission, we became aware of Ref.~\citen{KulishSklyanin}, in which the reductions of MNLSE equivalent to ours are derived.  The author thanks T. Tsuchida for informing him of this work. \\[0.5em]
\noindent\textbf{Acknowledgment}\\
	\indent This work was supported by a Grant-in-Aid for JSPS Fellows (No. 22-10058).

\end{document}